\documentclass[12pt,twoside]{article}
\usepackage{fleqn,espcrc1,epsf}
\title{The Infrared Behavior of QCD Propagators in
Landau Gauge\thanks{Supported by DFG (Al 279/3-2) and by the BMBF
(06-ER-809).}}

\author{Reinhard Alkofer$^a$
and Lorenz von Smekal$^b$\\
$^a$Institut f\"ur Theoretische Physik, Universit\"at T\"ubingen, 
Auf der Morgenstelle 14, D-72076 T\"ubingen, Germany. \\ 
$^b$Institut f\"ur Theoretische Physik III,
Universit\"at Erlangen--N\"urnberg,
Staudtstr.\ 7, D-91058 Erlangen, Germany.}

\begin{document}
\maketitle

\begin{abstract}
Some features of the solutions to the truncated Dyson-Schwinger
equations(DSEs) for the propagators of QCD in Landau gauge are summarized. In
particular, the Kugo-Ojima confinement criterion is realized, and positivity
of transverse gluons is manifestly violated in these solutions. In Landau
gauge, the gluon-ghost vertex function offers a convenient possibility to
define a nonperturbative running coupling. The infrared fixed point
obtained from this coupling which determines the 2-point interactions of
color-octet quark currents implies the existence of unphysical massless
states which are necessary to escape the cluster decomposition of
colored clusters. The gluon and ghost propagators, and the 
nonperturbative running coupling, are compared to recent lattice
simulations. A significant deviation of the running coupling from the
infrared behavior extracted in simulations of 3-point functions 
is attributed to an inconsistency of asymmetric subtraction schemes 
due to a consequence of the Kugo-Ojima criterion: infrared enhanced ghosts.  
\end{abstract}

\section{Confinement in Landau Gauge QCD}

Covariant quantum theories of gauge fields require indefinite
metric spaces. Modifications to the
standard (axiomatic) framework of quantum field theory are also necessary 
to accommodate confinement in QCD.
These seem to be given by the choice of either relaxing the principle of
locality or abandoning the positivity of the representation space. 
Great emphasis has therefore been put on the idea of relating confinement 
to the violation of positivity in QCD.  
Just as in QED, where the Gupta-Bleuler prescription is to
enforce the Lorentz condition on physical states, a semi-definite {\em
physical subspace} can be defined as the kernel of an operator.  
The physical states then correspond to equivalence classes of states 
in this subspace differing by zero norm components.  
Besides transverse photons covariance implies the existence of 
longitudinal and scalar photons in QED. The latter two form metric partners
in the indefinite space. The Lorentz condition eliminates
half of these leaving unpaired states of zero norm which do not contribute to
observables. Since the Lorentz condition commutes with the 
$S$-Matrix, physical states scatter into physical ones exclusively. 
Color confinement in QCD is ascribed to an analogous
mechanism: No colored states should be present in the positive definite space
of physical states defined by some suitable condition maintaining physical
$S$-matrix unitarity. Within the framework of BRS-algebra the completeness  
of the nilpotent BRS-charge $Q_B$ in a state space $\mathcal{V}$ of
indefinite metric is assumed. The (semi-definite) physical subspace 
$\mathcal{V_{\mbox{\tiny phys}}} = \mbox{Ker}\, Q_B$ is  
defined by those states which are annihilated by the BRS charge $Q_B$.
Positivity is then proved for physical states \cite{Kug79} which are 
given by the cohomology $\displaystyle \mathcal{H}(Q_B,\mathcal{V}) =
{\mbox{Ker}\, Q_B}/{\mbox{Im} Q_B} \simeq \mathcal{V}_s $, the covariant
space of equivalence classes of BRS-closed modulo exact states (in the image
${\mbox{Im} Q_B}$ of the BRS-charge).

\pagenumbering{arabic}
\stepcounter{page}
\markright{}

In perturbation theory the space $\mathcal{H}$ is formed by transverse gluon
and quark states. Longitudinal and timelike gluons form (massless) quartet
representations with ghosts and are thus unphysical. At the same time 
the global symmetry $J_{\mu ,\nu}^a$ corresponding to gauge
transformations generated by $\theta^a(x) = a^a_\mu x^\mu$ is spontaneously
broken quite analogous to the displacement symmetry in QED. 
Nonperturbatively, Kugo and Ojima have shown that the identification of
BRS-singlet states (in $\mathcal{V}_s$) with color singlets is generally
possible, in particular also for transverse gluons and quarks, if this global
symmetry is restored dynamically \cite{Kug79,Nak90,Kug95}. A sufficient 
condition for this restoration to happen is that the nonperturbative 
ghost propagator is more singular than a massless pole in the
infrared, {\it i.e.},   
\begin{equation}
    G(p) = \frac{-1}{p^2 ( 1 + u(p^2) )} \, , \;\; \mbox{with} \quad  
              u(p^2) \to -1 \;\; \mbox{for} \;\;  p^2 \to 0 \, .
\end{equation}
Furthermore, this mechanism is related to Nishijima's derivation \cite{Nis96}, 
from Ward-Takahashi identities, of the Oehme-Zimmermann
superconvergence relations which formalize a long known contradiction between
asymptotic freedom and the positivity of the spectral density for transverse
gluons in the covariant gauge \cite{Oeh80}. 

The remaining dynamical aspect of confinement in this
formulation resides in the cluster decomposition property of local field
theory. The proof of which, absolutely general other\-wise, does not
include the indefinite metric spaces of covariant gauge theories
\cite{Str76}. In fact, there  is quite convincing evidence for the contrary,
namely that the cluster property does not hold for colored correlations of
QCD in such a description \cite{Oji80}. This would thus eliminate the
possibility of scattering a physical state into color singlet states
consisting of widely separated colored clusters (the ``behind-the-moon''
problem, see also Ref.~\cite{Nak90} and references therein). Then, however,
there cannot be a mass gap in the whole indefinite space $\mathcal{V}$ 
(which implies nothing on the physical spectrum of the mass operator
in $\mathcal{H}$). 

Our solutions to the DSEs for the ghost and gluon propagators as reported in
Refs.~\cite{Sme97} provide for both, the Kugo-Ojima confinement criterion by
an infrared enhancement of the ghost propagator and the gapless spectrum by
the infrared fixed point in the color-octet correlations. The violation of
positivity of transverse gluons observed in~\cite{Sme97} seems unambiguously
established by a variety of independent nonperturbative studies of 
the gluon propagator as summarized in Ref.~\cite{Man99}. It is furthermore 
interesting to note that recent lattice calculations also 
verify the Kugo-Ojima criterion \cite{Nak99}.

\section{Infrared Behavior of Gluon and Ghost Propagators}

The known structures in the 3-point vertex functions can be employed to 
establish truncated DSEs that are complete for the gluon, ghost
and quark propagators of Landau gauge QCD \cite{Sme97,Sme98}. 
This is possible with systematically neglecting
contributions from explicit  4-point vertices to the propagator 
DSEs as well as non-trivial 4-point scattering kernels 
in the constructions of the 3-point vertices. 
The coupled DSEs for the propagators of the pure gauge theory can be solved
in a one-dimensional approximation \cite{Sme97}. 
Asymptotic expansions of the solutions in the infrared are obtained 
analytically. While the gluon propagator is found to vanish for small
spacelike momenta in this way, an apparent contradiction
with earlier studies that implied its infrared enhancement 
can be understood from the observation that the previously neglected ghost
propagator now assumes just this: An infrared enhancement of ghosts as
predicted by the Kugo-Ojima confinement criterion \cite{Kug95}. 
This infrared behavior of the propagators in Landau gauge 
was later confirmed qualitatively by studies of further truncated DSEs
\cite{Atk97}. Neither does it thus seem to depend on the particular 3-point
vertices nor on the one-dimensional approximation employed in our original
solutions. 

These solutions to the coupled gluon-ghost DSEs compare well with recent
lattice results available for the gluon \cite{Wil00,Cuc99} and the ghost
propagator \cite{Sum96} (which implement various lattice versions of the
Landau gauge condition). Indications towards an infrared
vanishing gluon propagator are now seen on the lattice also \cite{Cuc99}. At
least some suppression in the infrared is certainly excluded to be a 
finite size effect \cite{Wil00}. Especially the ghost propagator, however, is
in compelling agreement with the lattice data. This is an
interesting result for yet another reason: In \cite{Sum96,Cuc99} the Landau
gauge condition was supplemented by an algorithm to select gauge field
configurations from the fundamental modular region which is to avoid Gribov
copies. Thus, our results suggest that the existence of such gauge copies
might have little effect on the solutions to the Landau gauge DSEs. The
positivity violations of transverse gluons as seen in our results
\cite{Sme97} combined with the evidence from roughly a decade of lattice
simulations with considerably increasing statistics \cite{Ais97,Man99} leave
little doubt on the significance of this result. 

\section{Nonperturbative Running Couplings}
  
Lattice Landau gauge results have also become available for nonperturbative
running couplings from simulations of the 3-gluon \cite{All97,Bou98} and the
quark-gluon vertex \cite{Sku98}. These seem to have the common
feature of a maximum value $\alpha_S^{\hbox{\tiny{max}}}(\mu_0)$ at a {\em
finite} momentum scale $\mu_0 > 0$. Below this scale decreasing values of the
couplings are extracted towards smaller scales. Such qualitative forms lead
to double-valued $\beta$-functions, however. Here, we would like to point out
that this behavior seems likely to be an artifact of asymmetric
subtraction schemes in theories with confinement realized by the Kugo-Ojima
criterion. The reason is seen by relating results from asymmetric schemes
$\alpha_S^{\hbox{\tiny asym}}$ to those of symmetric subtraction schemes
$\alpha_S^{\hbox{\tiny sym}}$  which essentially results in a 
ratio of ghost renormalization functions 
\cite{Sme98},  
\begin{equation} 
\alpha_S^{\hbox{\tiny asym}}\!(\mu) \,  \propto  \, \lim_{s\to 0}
\frac{1+u(\mu^2)}{1+u(s)}  \,  \alpha_S^{\hbox{\tiny sym}}\!(\mu) \; . 
\end{equation}
With $u(0) = -1$ this is not well-defined. If the attention is limited to
the $\mu$-dependence of $\alpha_S^{\hbox{\tiny asym}}$, however, the
Kugo-Ojima criterion implies an artificial suppression in
$\alpha_S^{\hbox{\tiny asym}}\!(\mu)$ as compared to $\alpha_S^{\hbox{\tiny
sym}}\!(\mu)$. 

The (symmetric) scheme of \cite{Sme97} is based on the non-renormalization of
the ghost-gluon vertex in Landau gauge. The ideal comparison would thus be
provided by a lattice simulation employing this vertex to extract the running
coupling in a likewise symmetric subtraction scheme. It would be very
interesting to see whether such simulations could support the infrared fixed
point we obtained from the DSEs in \cite{Sme97} (with $\alpha_S(\mu \to 0)
\approx 9.5$ and monotonically decreasing for $\mu > 0$). This is because
such an infrared fixed point entails the existence of unphysical massless
(bound) states in the color-octet channels of 4-point functions (of
gluon/ghost and quark/antiquark correlations). It might also be worthwhile to
consider simulations of such 4-point functions in lattice Landau gauge, and
to assess possible indications for massless states in these correlations
directly. The existence of massless unphysical states is a necessary
condition for a failure of the cluster decomposition property for colored
clusters, see Chap.~4.3.4 in Ref.~\cite{Nak90}.

Recently we solved the coupled propagator DSEs with quarks included
\cite{Ahl00}. Implications on positivity violations and
confinement for quarks are currently being investigated.

\section*{Acknowledgements}
R.A.\ thanks the organizers of Quark Nuclear Physics 2000 for making this
stimulating conference possible. He is grateful to S.\ Furui, A.\ Schreiber,
A.\ Thomas and A.\ Williams for many helpful discussions. Furthermore, he
thanks all members of the CSSM for the extraordinary hospitality extended to
him at his stay after the conference.


\begin{thebibliography}{9}

\bibitem{Kug79}
T.~Kugo and I.\ Ojima, Prog. Theor. Phys. Supl. {\bf 66} (1979) 1.

\bibitem{Nak90}
M. Nakanishi and I.\ Ojima, ``Covariant Operator Formalism of Gauge Theories
and Gravity'', World Scientific Lecture Notes in Physics, Vol. 27, Singapore
1990. 

\bibitem{Kug95}
T.~Kugo, Int. Symp. on BRS symmetry, Kyoto, Sept.\ 18--22, 1995, 
hep-th/9511033.

\bibitem{Nis96}
K.~Nishijima, \newblock Czech.~J.~Phys. {\bf 46} (1996) 1; 
\newline
M.~Chaichian and K.~Nishijima, hep-th/9909158, and references therein.

\bibitem{Oeh80}
R.\ Oehme and W.\ Zimmermann, Phys. Rev. {\bf 21} (1980) 471, 1661.

\bibitem{Str76}
F.~Strocchi,
\newblock Phys.~Lett. {\bf B62} (1976) 60,
\newblock see also, Phys.~Rev.~{\bf D17} (1978) 2010.

\bibitem{Oji80}
I.~Ojima,
\newblock Z.~Phys. {\bf C5} (1980) 227.

\bibitem{Sme97}
L.~v.~Smekal, A.~Hauck and R.~Alkofer, Phys.\ Rev.\ Lett.\ {\bf 79} (1997), 
3591;\\  
L.~v.~Smekal, A.~Hauck and R.~Alkofer, Ann.\ Phys.\ {\bf 267} (1998), 1.

\bibitem{Man99}
J.~E.~Mandula, Phys. Rep. {\bf 315} (1999) 273.

\bibitem{Nak99}
S. Furui, these proceedings; H. Nakaijima and S. Furui, hep-lat/9909008.

\bibitem{Sme98}
R.~Alkofer, S.~Ahlig and L.~von~Smekal,
Fizika {\bf B8} (1999) 277 [hep-ph/9901322];\\
L.~v.~Smekal, habilitation thesis, Erlangen Univ., 1998, av. on request
from the author.

\bibitem{Atk97}
D.~Atkinson and J.~Bloch, Phys.\ Rev.\ \textbf{D58} (1998) 094036;\\
D.~Atkinson and J.~Bloch, Mod.~Phys.~Lett.~A \textbf{13} (1998), 1055.

\bibitem{Wil00}
A.G.~Williams, these proceedings;\\
 F.~D.~R.~Bonnet et al., hep-lat/0002020
and references therein.

\bibitem{Cuc99}
A.\ Cucchieri, hep-lat/9908050 and references therein.

\bibitem{Sum96}
H. Suman and K. Schilling, Phys. Lett. {\bf B373}, 314 (1996).

\bibitem{Ais97}
  H.~Aiso et al., Nucl.~Phys. B (Proc. Suppl.) {\bf 53} (1997), 570.

\bibitem{All97}
B.~All\'es et al., Nucl.~Phys.~B \textbf{502} (1997) 325.

\bibitem{Bou98}
Ph.~Boucaud et al.,  JHEP 10 (1998) 017; JHEP 12 (1998) 004. 

\bibitem{Sku98}
 J.~I.Skullerud, Nucl.~Phys.~Proc.~Suppl.\ \textbf{63} (1998) 242.

\bibitem{Ahl00}
S. Ahlig, R. Alkofer, C. Lerche, and L. von Smekal, in preparation.

\end{thebibliography}
\end{document}